УДК 004.056

# SECURITY AUDIT LOGGING IN MICROSERVICE-BASED SYSTEMS: SURVEY OF ARCHITECTURE PATTERNS

Alexander Barabanov[1], Denis Makrushin[2]

**Abstract**

**Objective**. Service-oriented architecture increases technical abilities for attacker to move laterally and maintain multiple pivot points inside of compromised environment. Microservice-based infrastructure brings more challenges for security architect related to internal event visibility and monitoring. Properly implemented logging and audit approach is a baseline for security operations and incident management. The aim of this study is to provide helpful resource to application and product security architects, software and operation engineers on existing architecture patterns to implement trustworthy logging and audit process in microservice-based environments.
**Method**. In this paper, we conduct information security threats modeling and a systematic review of major electronic databases and libraries, security standards and presentations at the major security conferences as well as architecture whitepapers of industry vendors with relevant products.
**Results and practical relevance**. In this work based on research papers and major security conferences presentations analysis, we identified industry best practices in logging audit patterns and its applicability depending on environment characteristic. We provided threat modeling for typical architecture pattern of logging system and identified 8 information security threats. We provided security threat mitigation and as a result of 11 high-level security requirements for audit logging system were identified. High-level security requirements can be used by application security architect in order to secure their products.

## 1 Introduction

Logging service in microservice-based systems is aim to meet principle of accountability and traceability and help to detect anomalies in operations via log analysis. Therefore, it is vital for applications security architects to understand and properly use existing architecture patterns to implement audit logging in microservices-based systems for security operations. The goal of our research was to identify such patterns and to do recommendations for applications security architect and security operations specialists on possible way to use it. This study is conducted with three main questions in mind:

- Threat modeling: what information security threats are exists for typical audit logging system in microservice-based applications?
- Security Design: what security control can be used while designing logging system to mitigate existing security threats?
- Implementation: What should application security architect take in mind while implementing audit logging system in microservice-based systems?

We provide threat modeling of simple logging system architecture pattern and reviewed major electronic databases and libraries (IEEE Xplorer, ACM Digital Library, SpringerLink, ResearchGate, arXiv) with research papers to extract primary studies. In order to explore these sources, we used search strings containing "logging", "audit", "monitoring", "log analysis", "security operations", "service-oriented architecture" and "microservice" (in different spelling, like "micro-service" or "micro service") words. To avoid missing relevant studies, we also reviewed security standards, presentations at the major security conferences and technical documents (whitepapaers) by industry vendors with mature microservice-based products.

[1] Alexander Barabanov, Ph.D, CISSP, CSSLP, Principal Security Engineer, Advanced Software Technology Laboratory, Huawei, Moscow, Russia. E-mail: barabanov.iu8@gmail.com
[2] Denis Makrushin, OSCP, Advanced Software Technology Laboratory, Head of Advanced Security Research Huawei, Moscow, Russia. E-mail:denis@makrushin.com

In summary, this paper makes the following contributions:

- threat model for typical logging system architecture pattern (Section 2);
- a set of security controls and mitigations techniques as well as non-functional requirements to the log format and set of auditable events (Section 3);
- recommendations for applications security architect on how to implement audit logging system in microservice-based applications (Section 4).

This article continues a set of articles dedicated to microservice-based system security [14].

## 2 Threat Model for Audit Logging System

We provided security design review in order to define typical security threats and mitigation techniques. To define security threats we used STRIDE methodology [1], [3], CAPEC [2] repository and best practices analysis. During best practices analysis we analyzed architectural patterns used in the wild and presented at application security conferences[3][4][5].A naive pattern of logging subsystem is shown on the picture below (Figure 1). Microservice directly sends their log message to central logging service via network requests using logging library (e.g., log4j for Java-based applications).

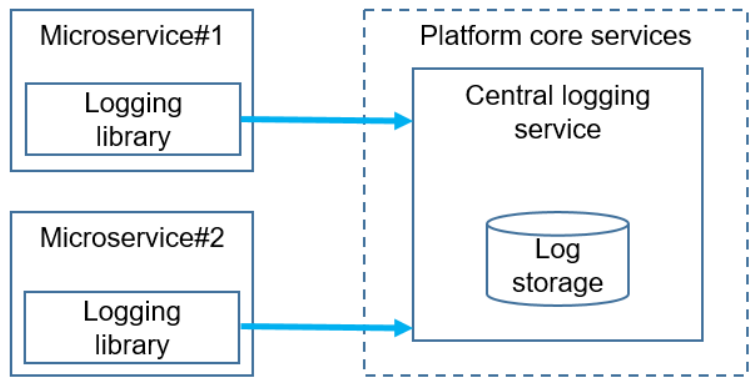

Figure 1 Logging pattern "Microservice directly sends log message to central logging"

We identify 8 threat categories against audit logging subsystems (Table 1) for the mentioned naïve pattern of logging subsystem. It should be mention, that our security design review was scoped only to collection layer and transport/streaming layer [4]. Other layers (analysis, storage and access) was out of the current research scope (e.g. security threats to log storage integrity).

Table 1

| ID | Threat definition | CAPEC reference (if applicable) |
|---|---|---|
| Th#1 | Service spoofing: a malicious or compromised microservice can send log message to the logging subsystem to forge logs | CAPEC-151: Identity Spoofing |
| Th#2 | Logging/transport system spoofing: a malicious service can act as a logging subsystem in order to get access to sensitive information | CAPEC-151: Identity Spoofing |

---

[3] P. Phadnis, S.Nagmote. (2019). Massive Scale Data Processing at Netflix using Flink. Talk presented at the Flink Forward 2019
[4] K. Gade, Yu Yang. (2016). Scalable and Reliable Logging at Pinterest. Talk presented at the DataEngConf SF16
[5] M.Koes (2018). Centralized Logging Solution for Google Cloud Platform. Talk presented at the Cloud Next '18

| Th#3 | Logging system denial-of-service (DoS): data loss due to logging service failure in case of attack on logging service. In that case log messages needed to be buffered on the microservice side in memory. As the microservice buffer size is limited, an extended logging service outage would lead to log message loss | CAPEC-125: Flooding |
|---|---|---|
| Th#4 | An adversary modifies content (log message published by microservice to logging subsystem via communication channel) to make it contain something other than what the original content producer (microservice) intended while keeping the apparent source of the content unchanged | CAPEC-594: Traffic Injection |
| Th#5 | Adversary intercepts information transmitted between microservice and logging subsystems to capture sensitive information | CAPEC-158: Sniffing Network Traffic |
| Th#6 | Legitimate microservice (due to attack) can elevate its privileges in order to read sensitive information from logging service | CAPEC-122: Privilege Abuse |
| Th#7 | Data loss due to logging service failure in case of its flooding by legitimate microservice (highload environment). In that case log messages needed to be buffered on the microservice side in memory. As the microservice buffer size is limited, an extended logging service outage would lead to log message loss. | CAPEC-130: Excessive Allocation |
| Th#8 | Microservice may log private or confidential data (e.g., PII, passwords, API keys) without masking/filtering. An attack to the logging service may lead to sensitive information disclosure. | CAPEC-215: Fuzzing and observing application log data/errors for application mapping |

## 3 Audit logging system security controls and requirements

Then we analyzed identified security threats as well as best practices [4], [6], [7], [8], [9] adopted by community and industry vendors with mature product security program in order to identify high-level requirement for logging subsystem. A high-level architecture design is shown on the picture below ( Figure *2*):

- microservice writes a log messages to local file using standard output (via stdout, stderr);
- logging agent periodically pulls log messages and sends (publish) it to message broker;
- central logging service subscribes on messages in message broker, receives and process it.

Logging agent can be deployed using daemonset in Kubernetes environment[6] or sidecar pattern[7].

---

[6]A Kubernetes DaemonSet is a container tool that ensures that all nodes or a specific subset of them are running exactly one copy of a pod

[7] See information about "sidecar" pattern on https://microservices.io/patterns/deployment/sidecar.html

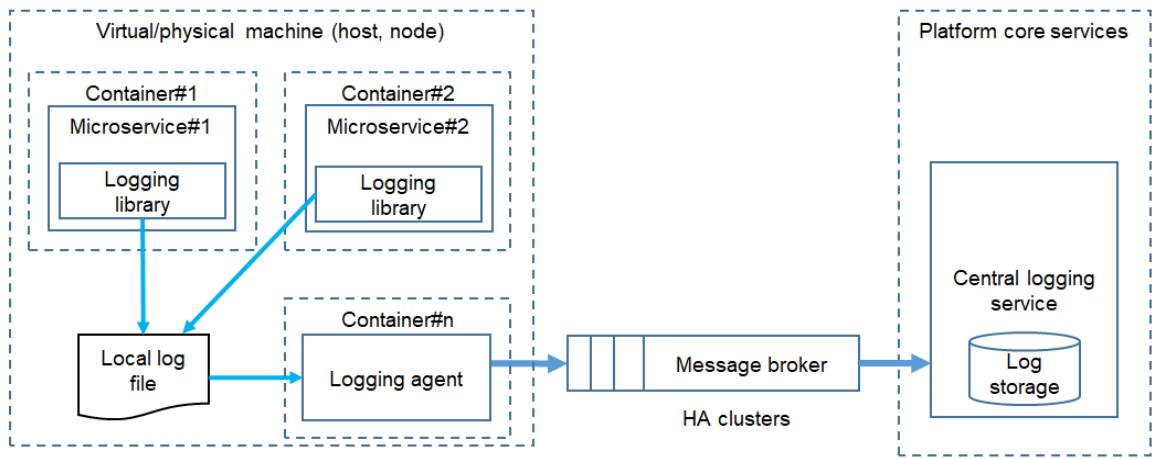

Figure 2 Logging pattern "Logging agent"

High-level requirements to logging subsystem architecture with its rationales are listed in table below (Table 2).

Table 2

| High-level requirement definition | Rational |
|---|---|
| Microservice shall not send log message directly to central logging subsystem using network communication. Microservice shall write its log message to a local log file. | Mitigated threats:<br>• Th#3 Data loss due to logging service failure (due to attack): in case of logging service outage microservice will still write log messages to the local file (without data loss), after logging service recovery logs will be available to shipping<br>• Th#7   Data loss due to logging service failure in case of its flooding by legitimate microservice (highload environment): in case of logging service outage microservice will still write log messages to the local file (without data loss), after logging service recovery logs will be available to shipping |
| There shall be a dedicated component (logging agent) decoupled from microservice. Logging agent shall collect log data on microservice side (read local log file) and send it to central logging subsystem. Due to possible network latency issues logging agent shall be deployed on the same host (virtual or physical machine) with microservice. | Mitigated threats:<br>• Th#3 Data loss due to logging service failure (due to attack): isolate microservice from logging agent failure - in case of logging agent failure microservice still writes information to the log file, logging agent after recovery will read file and send information to message broker; |
| Due to possible DoS-attack on central logging subsystem logging agent shall not uses synchronous request/response pattern to send log messages. There shall be message broker to implement asynchronous connection between logging agent and central logging service. | Mitigated threats:<br>• Th#7   Data loss due to logging service failure in case of its flooding by legitimate microservice (highload environment): using asynchronous mechanism allows to mitigate possible DoS-attack from legitimate microservices because microservice process amount of information it can that do not lead to exhaustion of microservice resources |
| Logging agent and message broker shall use mutual authentication (e.g. based on TLS) to encrypt all transmitted (log messages) and authenticate themselves. | Mitigated threats:<br>• Th#1 Microservice spoofing<br>• Th#2 Logging/transport system spoofing<br>• Th#4 Network traffic injection<br>• Th#5 Sniffing Network Traffic |
| Message broker shall enforce access control policy in order to mitigate unauthorized access and enforce the principle of least privileges | Mitigated threats:<br>• Th#6 Microservice elevation of privileges |
| Logging agent shall filter/sanitize output log messages in order to sensitive data (e.g., PII, passwords, API keys) will never send to the central logging subsystem (data minimization principle) | Mitigated threats:<br>• Th#8 Microservice may log private or confidential data without masking/filtering |
| Message broker shall be deployed in high availability mode (cluster) | Mitigated threats:<br>• Th#7 Data loss due to logging service failure in case of its flooding by legitimate microservice (highload environment) |
| Microservices shall generate and pass through microservice call chain a correlation ID which uniquely identify every call chain and help grouping log messages to investigate them.<br>Logging agent shall include correlation ID in every log message. | Best practices [4] |
| Logging agent shall periodically provide health and status data to indicate its availability or non-availability | Best practices [4] |
| Logging agent shall publish log messages in structured logs format (e.g., JSON, CSV) | Best practices [4] |
| Logging agent shall append log messages with context data (e.g., platform context, runtime context) | Best practices [8][9] [4] |

We analyzed several research papers, books and presentations [4][10], to identify a set of events that should be logged. Possible auditable events at microservice/application level are presented in the table below (Table 3).

Table 3

| Auditable events | Notes |
|---|---|
| Authentication successes and failures | • Incorrect username/password<br>• Token (e.g., JWT) validation issues (incorrect signature, incorrect signature method, absence of JWT) |

---

[8] Chandramouli R. (2019) Security Strategies for Microservices-based Application Systems. (National Institute of Standards and Technology, Gaithersburg, MD), NIST Special Publication (SP) 800-204. https://doi.org/10.6028/NIST.SP.800-204

[9] Chandramouli R., Butcher Z. (2020) Building Secure Microservices-based Applications Using Service-Mesh Architecture. (National Institute of Standards and Technology, Gaithersburg, MD), NIST Special Publication (SP) 800-204A. https://doi.org/10.6028/NIST.SP.800-204A

[10] A.Fontaine (2018). Logging in the age of Microservices and the Cloud. Talk presented at DevOps Con 2018

| Authorization (access control) failures | Every access control decision should be logged. Different access control framework implement built-in logging, e.g. see example of Open Policy Agent authorization decision log on the picture below. |
|---|---|
| Application errors and system events | <ul><li>syntax and runtime errors;</li><li>connectivity problems;</li><li>performance issues, third party service error messages;</li></ul> |
| Use of higher-risk functionality | <ul><li>network connections (other microservice request);</li><li>addition or deletion of users;</li><li>changes to user privileges;</li></ul> |
| Input validation failures | <ul><li>protocol violations</li><li>unacceptable encodings</li><li>invalid parameter names and values</li></ul> |
| Application/microservice states changes | <ul><li>start-ups</li><li>shut-downs</li><li>logging initializations</li><li>logging configuration updates</li></ul> |

```
[
  {
    "labels": {
      "app": "my-example-app",
      "id": "1780d507-aea2-45cc-ae50-fa153c8e4a5a",
      "version": "latest"
    },
    "decision_id": "4ca636c1-55e4-417a-b1d8-4aceb67960d1",
    "bundles": {
      "authz": {
        "revision": "W3sibCI6InN5cy9jYXRhbG9nIiwicyI6NDA3MX1d"
      }
    },
    "path": "http/example/authz/allow",
    "input": {
      "method": "GET",
      "path": "/salary/bob"
    },
    "result": "true",
    "requested_by": "[::1]:59943",
    "timestamp": "2018-01-01T00:00:00.000000Z"
  }
]
```

Figure 3 Access control decision log example (Open Policy Agent)

Using structured logs format (e.g., JSON, CSV) is a widely used pattern. It is advisable to define a structure for log events among development teams. The following example (Figure 4) shows log message in mozlog format [4].

```
{
    "Timestamp": 145767775123456,
    "Type": "request.summary",
    "Logger": "myapp",
    "Hostname": "server-a123.mozilla.org",
    "EnvVersion":"2.0",
    "Severity": 6,
    "Pid": 1337,
    "Fields":{
```

```
            "agent": "curl/7.43.0",
            "errno": 0,
            "method": "GET",
            "msg": "the user wanted something.",
            "path": "/something",
            "t": 5,
            "uid": "12345"
        }
}
```

Figure 4 Log example (mozlog library)

Possible log message fields are listed in the table below (Table 4).

Table 4

| Category | Fields/event attribute |
|---|---|
| Time | May be:<br>• consistent ISO8601 dates with nanosecond precision;<br>• number of nanoseconds since the UNIX epoch;<br>• field that complies to RFC3339 |
| Level | The level of the log: INFO, DEBUG, ERORR |
| Message | Human readable string |
| HTTP context:<br>context about the HTTP request currently being processed (if any) | • host<br>• user agent string<br>• HTTP method<br>• path<br>• remote address<br>• correlation ID<br>• user ID |
| Platform context:<br>e.g. contextual information about the Pod running on the Kubernetes platform | • host<br>• pod name<br>• pod ID<br>• container name<br>• namespace name<br>• namespace ID |
| Runtime context:<br>represents the language runtime details | • class name<br>• file name<br>• function<br>• module name |
| Source context: context about the source of the log | • IP address<br>• PID<br>• process name<br>• user ID |
| Custom data | Microservice can add any valuable structured data |

## 4. Recommendations for application security architects

Based on our survey results, we came up with several recommendations for application security architects on audit logging implementation:
1) Centralized logging service allow to collect, stream, analyze, store, and access log events. Using a centralized logging service that aggregates logs from each service instance is a widely used pattern. To mitigate typical threats related with logging system it is advisable to use logging agents and publish-subscribe pattern.
2) Audit logging solution should be infrastructure/operations-level solution and dedicated team (e.g., Infrastructure security team) must be accountable for development and its operation as well as sharing microservice blueprint/library/components that implement audit logging among development teams. Audit logging solution should be easy to implement for development teams and scalable. It is preferable to ask development teams to use standard output to write log messages.

3) Application security architect should establish a baseline related with auditable events, log format and log message fields and share it among development and operations teams. Auditable events list should include: authentication successes and failures, authorization (access control) failures, application errors and system events, use of higher-risk functionality, input validation failures, application/microservice start-ups and shut-downs, logging initialization. Using structured logs format (e.g., JSON, CSV) is a widely used pattern.

4) Audit logging system should be based on widely used solution, because implementing custom solution has following cons:
   - security or engineering team have to build and maintain custom solution;
   - it is necessary to build and maintain client library SDKs for every language used in system architecture;
   - necessity to train every developer on custom audit API and integration, and there's no open source community to source information from.

## 5. Related work

Security architecture patterns for microservice-based systems has been the topic of a number of surveys and review articles, as well as standards.

Vale et al. [10] conducted a systematic mapping to reveal adopted security mechanisms for microservice-based systems. They focused only on security mechanisms and examined 26 papers published from November 2018 to March 2019. Hannousse et al. [11] conducted a similar investigation to Vale et al. [10] study. Their study is broader in several ways: they included published papers since 2011 and besides security mechanisms, they also focused on identifying security threats and the applicability of proposed solutions regarding their execution platforms and architectural layers. Yu et al. [12] surveyed work related to security risks for microservices-based fog applications, and argued that security issues arise in four system aspects: containers, data, permissions and network security. Nehme et al. [13] discussed how microservices can be secured at different levels and stages considering a common software development lifecycle.

NIST published standards[11][12] on microservice-based system security. NIST analyzed the multiple implementation options available for each individual core security feature (authentication and access management, service discovery, secure communication protocols, security monitoring, availability/resiliency improvement techniques, load balancing and throttling, integrity assurance techniques and handling of session persistence) and configuration options in architectural frameworks, and developed security strategies that counter threats specific to microservice-based systems.

Compared with the related works our study is more narrow and concentrated on audit logging function only in order to get deeper results. Moreover, besides research papers analysis we also analyzed presentations at the major security conferences.

---

[11] Chandramouli R. (2019) Security Strategies for Microservices-based Application Systems. (National Institute of Standards and Technology, Gaithersburg, MD), NIST Special Publication (SP) 800-204. https://doi.org/10.6028/NIST.SP.800-204

[12] Chandramouli R., Butcher Z. (2020) Building Secure Microservices-based Applications Using Service-Mesh Architecture. (National Institute of Standards and Technology, Gaithersburg, MD), NIST Special Publication (SP) 800-204A. https://doi.org/10.6028/NIST.SP.800-204A

# 6. Conclusion and further work

The survey enumerated several technical publications and multiple sources and libraries to deliver a source of insight for application security architects to build trustworthy audit logging system. This document offers the benefit of providing guidance on needed efforts and research into how to better secure microservice-based environments.

Security operation team should actively uses logged information to detect security threats. In order to do that security analyst can use simple patterns based on regular expressions and statistics and threat detection based machine learning techniques [15], [16]. The work can be a baseline for future audit logs processing by establishing an efficient hybrid approach that combines classic techniques to detect threats in microservice-based environment and novel approaches (e.g., machine-learning based algorithms) to detect anomalies and adversaries in compromised infrastructure.

# АУДИТ СОБЫТИЙ БЕЗОПАСНОСТИ В МИКРОСЕРВИСНЫХ ПРИЛОЖЕНИЯХ: ОБЗОР АРХИТЕКТУРНЫХ ПОДХОДОВ

Барабанов А.[13], Макрушин Д.[14]

**Аннотация**

**Цель статьи**. Использование сервис-ориентированной архитектуры при проектировании программного обеспечения открывает новые возможности для нарушителей, которые используют новые методы закрепления и перемещения внутри скомпрометированной инфраструктуры.Микросервисы приносят новые задачи для архитекторов безопасности, связанные с повышением уровня мониторинга событий внутри защищаемой среды. Целью данного исследования является создание базы типовых архитектурных решений, которые могут быть использованы разработчиками и архитекторами информационной безопасности при проектировании и реализации функций сбора и аудита событий внутри инфраструктуры, основанной микросервисах.

**Метод исследования** заключается в моделировании угроз и системном анализе научных публикаций и выступлений на ведущих научно-технических конференциях по теме защиты информации в микросервисных приложениях, обобщении и систематизации полученных результатов.

**Полученные результаты и практическая значимость**. В работе представлен систематизированный перечень архитектурных подходов, которые могут быть использованы для реализации системы сбора и аудита событий в микросерсивных приложениях. Представлены результаты моделирования угроз безопасности информации (идентифицировано 8 угроз), выполненного в отношении типового архитектурного шаблона реализации системы системы сбора и аудита событий. По результатам анализа идентифицированных угроз безопасности информации разработаны 11 требований к обеспечению безопасности систем сбора и аудита событий в микросервисных приложениях. Разработанные требования могут быть использованы архитекторами информационной безопасности при проектировании и реализации микросервисных приложений.

**Ключевые слова:** микросервис, регистрация событий, архитектурные подходы, защита информации

## Литература

---

[13] Александр Барабанов, канд.техн.наук, CISSP, CSSLP, Лаборатория передовых программных технологий, ведущий научный сотрудник, компания Huawei, Москва, Россия. E-mail: barabanov.iu8@gmail.com

[14] Денис Макрушин, OSCP, Лаборатория передовых программных технологий, руководитель направления перспективных исследований безопасности, компания Huawei, Москва, Россия. E-mail:denis@makrushin.com